\documentclass[english,floatfix,twocolumn,showpacs,superscriptaddress,nofootinbib,aps,prd]{revtex4}
\usepackage[T1]{fontenc}
\usepackage{lmodern}
\usepackage{amsmath}
\usepackage{amssymb}
\usepackage{graphicx}
\usepackage{esint}
\usepackage{longtable}
\usepackage{dcolumn}
\usepackage{babel} 
\usepackage{csquotes}

\begin{document}

\title{Accretion of perfect fluids onto a class of regular black holes}

\author{Juliano C. S. Neves}
\email{nevesjcs@if.usp.br}
\affiliation{Centro de Ciências Naturais e Humanas, Universidade Federal do ABC,\\ Avenida dos Estados 5001, Santo André, 09210-580 São Paulo, Brazil}


\author{Alberto Saa}
\email{asaa@ime.unicamp.br}
\affiliation{Departamento de Matemática Aplicada, Universidade
Estadual de Campinas, \\
 13083-859 Campinas-SP, Brazil}

\begin{abstract}
We consider the stationary spherical  accretion process of perfect fluids onto a class of spherically
symmetric  regular black holes corresponding to quantum-corrected Schwarzschild spacetimes. 
We show that the accretion rates can differ from  the Schwarzschild case,  suggesting that the de Sitter 
core inside these regular black holes, which indeed precludes the central singularity, can act  for some cases  as 
a sort of antigravitational source, decreasing the fluid's radial infall velocity in the accretion process, and for 
others  as a gravitational enhancer, increasing the fluid flow into the black hole horizon.  Our analysis and results 
can be extended and also applied to the problem of  black hole evaporation in cosmological scenarios with 
phantom fluids. In particular, we show that the mass of typical regular black holes can be used in order to constrain 
turnaround events in cyclic cosmologies.  
\end{abstract}


\maketitle

\section{Introduction}

A Regular black hole  (RBH) is a solution of the gravitational field equations without a singularity inside the event
horizon. The first RBH ever created is due to  Bardeen  \cite{Bardeen}. It   possesses spherical
symmetry and a mass function that depends on the radial coordinate. Inside the event horizon, the Bardeen RBH hides a de
 Sitter core which precludes   either the point singularity of
 spherical RBHs \cite{Ansoldi,Lemos_Zanchin,Dymnikova,Dymnikova2,Dymnikova3,Bronnikov,Hayward,Neves,Neves2} or the ring singularity in RBHs with axial symmetry \cite{Various_axial,Various_axial2,Various_axial3,Various_axial4,Various_axial5,Neves_Saa}. As it is
 well known, the de
 Sitter core promotes energy conditions violations in order to avoid the conclusions of the singularities theorems (see
 \cite{Wald} for a detailed study on the theorems). For example, the Bardeen RBH violates the strong energy
 condition, and RBHs with rotation ignores the weak energy condition \cite{Neves_Saa}.   

The class of RBHs studied in this work adopts a generalized mass function that includes the Bardeen \cite{Bardeen} and Hayward \cite{Hayward} mass functions. The mass function is
\begin{equation}
m(r)=M_{0}\left(1+\left(\frac{r_{0}}{r}\right)^{q}\right)^{-\frac{3}{q}},
\label{Mass_term}
\end{equation}
and it
provides RBHs solutions for any positive integer $q$, by assuming  geometries with either spherical or axial 
symmetry. For example, considering spherical symmetry, for $q=2$ one has the so-called Bardeen RBH, and when $q=3$ the
interesting Hayward RBH is obtained. However, other RBHs are possible for different values of $q$
or different integers (for example,
$q=1$ produces a RBH and will be also studied here), 
they are new geometries and are asymptotically Schwarzschild for large values
of the radial coordinate, outside the event horizon. 
Notice that, as discussed in \cite{Neves_Saa}, 
it is 
 possible to interpret the interior region of these RBHs, for any value of $q$, as a de Sitter solution  with high matter-energy densities,
in the same spirit of the pioneering works of  
 Sakharov \cite{Sakharov} and Gliner \cite{Gliner}.
The parameter $M_0$ is the Arnowitt-Deser-Misner (ADM) mass for the spherical and asymptotically flat case.
The length parameter 
 $r_{0}$ has been recently   
  investigated \cite{Maluf_Neves2} and, by 
    using a generalized uncertainty principle (GUP), 
a   new interpretation of the Bardeen metric was proposed: 
the Bardeen RBH can be conceived of as a quantum-corrected Schwarzschild black hole instead of a BH that comes from a nonlinear electrodynamics as is well-known in Ayón-Beato and Garcia's work \cite{Beato}. This new interpretation led to an upper bound on the length parameter, namely $r_0<10^{-25}$m, which is
compatible with a microscopical origin. As we will see, by using this upper bound, we will calculate important quantities in the accretion process like the radial velocity and the critical point where the fluid's velocity equals the critical speed.  

The accretion process onto BHs is a remarkable issue in science today. The recent image of a BH constructed 
by the Event Horizon Telescope \cite{EHT} confirms that. The accretion of perfect fluids onto BHs,
in particular, is also an issue with a
large literature. In a seminal paper in the 70s, which is a development of Bondi's 
work \cite{Bondi} 
on accretion in the Newtonian context, Michel \cite{Michel} investigated the steady-state accretion of a 
perfect fluid onto the Schwarzschild BH. A steady-state accretion process does not depend on time, it is a stationary
process. For an observer in the spherically symmetrical case, physical quantities in such a process are $r$-dependent only. Following Michel, Babichev \textit{et
 al}. \cite{Babichev1} studied the accretion of a phantom fluid onto the Schwarzschild metric. (For
 the analysis involving scalar fields instead of fluids, see \cite{RodriguesSaa} and references
 therein.) As a continuation, the same
authors \cite{Babichev2} used the Reissner-Nordström geometry in order to obtain the accretion equations (in the same
 direction, see \cite{Mubasher} for a study on accretion onto charged BHs). Cases with a cosmological constant were
 studied in \cite{Abbas,Amani}, accretion onto general spherical BHs were discussed in \cite{Bahamonde,Chaverra}, and
  the same process for scale-dependent black boles is presented in \cite{Contreras}. In the first work on accretion onto a
   RBH \cite{Abkar}, the Bardeen   metric was used. Studies on accretion onto other RBHs are shown in
    \cite{Debnath,Jawad}. In the  present work, we investigate the accretion process of perfect fluids for RBHs generated by the mass function (\ref{Mass_term}) and, then, compare to the Schwarzschild case.

As pointed out by Babichev \textit{et al.} \cite{Babichev1}, 
it is also worth considering the 
 phantom fluid case because such a fluid decreases the BH mass in the accretion process. A perfect fluid
 is said to be  phantom-like when its equation of state (EoS) is $w=\frac{p}{\rho}<-1$ or, equivalently, $p+\rho<0$, where $p$ and $\rho$ are the fluid's pressure and energy density, respectively. Recently, the Planck Collaboration \cite{Planck}, by using Planck + Pantheon supernovae + BAO data, constrained the dark energy EoS to $w=-1.03\pm 0.03$. Some years before, the Supernova Cosmology Project obtained $w=-0.997^{+0.077}_{-0.082}$ for a flat universe \cite{Supernova}.  With such an EoS, a phantom fluid that describes the dark energy would be pretty much ruled out. However, according to the authors of Ref. \cite{Shafer}, who adopted the Union2 data from Supernova Project, the phantom energy could be a candidate for dark energy with $\sim1.9\sigma$. For that reason, disintegration of BHs by phantom-like fluids is an important issue in the cosmological scenario as well. In bouncing cosmologies or cyclic models \cite{Novello,Brandenberger1}, which are options to the standard big bang model, elimination of BHs is essential in order to promote models beyond the standard cosmology \cite{Neves3,Brandenberger2,Frampton,Brown,Sun}. A given cycle cannot inherit large BHs from the previous cycle. The existence of cyclic models depends on eliminating BHs with the aid of a phantom energy or Hawking's radiation \cite{Frampton,Brown,Sun}. With the turnaround critical mass, obtained by Ref. \cite{Sun} within a cyclic scenario, we compare the Schwarzschild critical mass to the critical mass of the class of RBHs studied here. As we will see, there will be a slight difference over time.   

The structure of this paper is as follows: in Section II we derived the equations that describe the steady-state 
accretion process for spherical RBHs with mas term (\ref{Mass_term}), and we compared the accretion process 
regarding the Schwarzschild BH and RBHs; in Section III some types of fluids and critical points are evaluated. 
In Section IV, implications on cosmology are briefly discussed, and the final remarks are made in Section V.
In this work, we adopted geometric units such that $G=c=1$, where $G$ is the gravitational constant, and $c$ is the speed of light in vacuum.

\section{The steady-state accretion process}

Michel's approach \cite{Michel} assumes some important conditions in order to provide the accretion equations for static spacetimes:
\begin{enumerate}
\item General Relativity is the theory of gravitation and, thus, the matter energy-momentum tensor is conserved;
\item the accretion  is a steady-state process with the accreted matter   conceived of as a perfect fluid;
\item  the fluid in the process does not modify the original geometry, {\em  i.e.}, back-reaction effects due to the fluid's self-gravity are neglected.
\end{enumerate}
Michel's approach 
gives origin to both   physical and nonphysical solutions. We are interested in physical ones, {\em i.e.}, those ones corresponding to \enquote{real} accretion  phenomenon, for which the perfect
fluid    departs from infinity at rest and
 has its infall velocity monotonically 
 increased in the outside region of the black hole event horizon. 
 For these solutions, the fluid experiences 
 a subsonic flow in the regions far from the horizon, and then crosses a critical point ($r=r_c$), also outside the event horizon. At the critical point, the sonic point, the fluid has radial velocity equal to the speed of sound, and after that point in the inflow motion the fluid's radial velocity increases to the supersonic levels. Therefore, regarding spherical spacetimes, we have a sonic sphere at $r=r_c$, in which every fluid's element has radial velocity equal to speed of sound. 

Let us obtain the process equations for static metrics. First of all, it is assumed a  metric with spherical symmetry in the $(t,r,\theta,\phi)$ coordinates in the form
\begin{equation}
ds^{2}=-f(r)dt^{2}+\frac{dr^{2}}{f(r)}+r^{2}\left(d\theta^{2}+\sin^{2}\theta d\phi^{2}\right),
\label{Metric}
\end{equation}
where $f(r)=1-2m(r)/r$, and the mass function is given by Eq. (\ref{Mass_term}). Due to the shape of $m(r)$,
the spacetime structure of (\ref{Metric}) is quite different from the Schwarzschild geometry. That is, when $r_{0}\ll M_{0}$ the geometry given by $m(r)$ has two horizons, namely an inner $(r_{-})$ and an outer horizon, the event horizon ($r_{+}$). In the case when $r_{0}>M_{0}$, there are no horizons (this is not the case studied here). In the latter case, neither horizons nor naked singularities are present because the mass function $m(r)$ avoids a singularity, and $f(r)$ does not provide any roots. 

A perfect fluid has an energy-momentum tensor given by 
\begin{equation}
T_{\mu\nu}=(p+\rho)u_{\mu}u_{\nu}+pg_{\mu\nu},
\label{Energy_momentum}
\end{equation}
where $p$ and $\rho$ are the proper pressure and proper energy density, respectively, and obey an EoS $p(\rho)$. The four-vector $u^\mu$ is the four-velocity of the fluid and, in the radial process, is simply $u^{\mu}=(u^t,u^r=u,0,0)$, 
with $u^{\mu}=dx^{\mu}/ds$ and the following normalization: $u^{\mu}u_{\mu}=-1$. Following Michel, 
the first equation of accretion is obtained from the current conservation, $J_{;\mu}^{\mu}=0$, which indicates 
the conservation of the mass flux (a semicolon indicates a covariant derivative). As the current density is
given by $J^{\mu}=m_0n u^\mu$, where $m_0$ is the mass of each particle in the fluid and $n$ is 
the proper particle number density, then the current conservation leads to\footnote{The major part of the quoted works \cite{Mubasher,Bahamonde,Jawad} adopts $\rho$ instead of $n$ in Eq. (\ref{C1}). However, this is true just for a pressureless fluid.}
\begin{equation}
n ur^{2}=C_{1}.
\label{C1}
\end{equation}
The second equation for the accretion process  comes from the energy-momentum conservation, $T_{0;\mu}^{\mu}=0$. By using $u^{\mu}u_{\mu}=-1$ and $u^t=\sqrt{\frac{1}{f(r)}+\frac{u^2}{f(r)^2}}$, the second equation can be written as 
\begin{equation}
\left(p+\rho\right)\left(f(r)+u^{2}\right)^{\frac{1}{2}}ur^{2}=C_{2}.
\label{C2}
\end{equation}
  As we pointed out, the component $u$ is the radial velocity  $u=\frac{dr}{ds}$, and $u<0$ for accretion, {\em i.e.}, for the fluid's inflow motion. With the aid of (\ref{C1}), Eq. (\ref{C2}) is rewritten as
\begin{equation}
\left(\frac{p+\rho}{n}\right)^{2}\left(f(r)+u^{2}\right)=C_{3}=\left(\frac{C_{2}}{C_{1}}\right)^{2}.
\label{C3}
\end{equation}
It is worth mentioning some important results and limits that are obtained from Eq. (\ref{C3}). Assuming a
 dust fluid ($p=0$, thus $n \varpropto \rho$), alongside the condition of rest at
infinity, namely $u_{\infty}\rightarrow 0$, we will have $C_{3}=1$ and the Newtonian equation for freely falling bodies,
that is to say, $u_{Sch}=-\sqrt{2M_0/r}$ for each fluid's particle, if the metric is the Schwarzschild metric. 
On the other hand, an interesting result  is obtained from Eq. (\ref{C2}) for a cosmological constant, 
$p=-\rho$, in the accretion process. 
As we can see, Eq. (\ref{C2}) indicates lack of accretion if the cosmological constant is conceived of as a perfect fluid 
with $w=-1$. As pointed out in
 \cite{Ademir}, a Schwarzschild-(a)-dS BH does not provide an accretion process of a cosmological constant as well.    

Indeed, regarding the Schwarzschild metric and the class of RBHs generated by the general mass function given by Eq. (\ref{Mass_term}),  it is possible to compare the radial velocity for the entire class of RBHs with the radial velocity of the accretion process onto the Schwarzschild BH, by using an  approximation for the mass function (\ref{Mass_term}). For that purpose, we write the mass function as 
\begin{equation}
m_l(r) \simeq M_{0}\left(1-\frac{3}{q}\left(\frac{r_{0}}{r}\right)^{q}\right),
\label{Mass_approx_large}
\end{equation}
for large values of the radial coordinate. It is worth emphasizing that for small $r$, it is problematic to speak of radial velocities due to the absence of a stationary observer inside the event horizon. On the other hand, for large values of $r$, assuming again the EoS $w=\frac{p}{\rho}=0$ and $n \varpropto \rho$, Eq. (\ref{C3}) presents the following result by adopting (\ref{Mass_approx_large}):
\begin{equation}
u_l = -\sqrt{\frac{2M_0}{r}\left(1-\frac{3}{q}\left(\frac{r_0}{r} \right)^q \right)}.
\label{u_large}
\end{equation}
 A straightforward interpretation indicates that the fluid's radial velocity (for dust fluid)  is smaller  for the entire 
 class of RBHs compared to the Schwarzschild accretion velocity. Of course, due to the very small size of   $r_0$, 
 the difference is very tiny in Eq. (\ref{u_large}) for astrophysical objects, whether stellar or supermassive BHs. Nevertheless, we believe that the tiny difference between the class of RBHs and the Schwarzschild BH comes from  
 the de Sitter core inside RBHs. In this case, the de Sitter core works as an antigravity source, 
decreasing the radial velocity in the accretion process onto RBHs.\footnote{As we will see, this is not the case
for other fluids.} This core's capability was also indicated in geodesic
 studies of the Bardeen RBH in Ref. \cite{Stuchlik}. As we can see by adopting $m(r) \simeq M_0(r/r_0)^3$ for the mass function when $r$ is small, the metric term is $f(r)\sim 1-Cr^2$ for those values of $r$, with $C$ playing the role of a positive cosmological constant and illustrating the de Sitter spacetime inside the RBHs of the class studied here.      

Differentiating (\ref{C1}) and (\ref{C3}) with elimination of $dn$ provides 
\begin{equation}
\frac{du}{dr}=-\frac{u}{r}\frac{\left(2V^{2}\left(f(r)+u^{2}\right)-\frac{1}{2}rf'(r)\right)}{\left(V^{2}\left(f(r)+u^{2}\right)-u^{2}\right)},
\label{Critical1}
\end{equation}
where
\begin{equation}
V^{2}\equiv\frac{d\ln(p+\rho)}{d\ln n}-1,
\label{V}
\end{equation}
with the symbol $(')$ denoting differentiation with respect to the radial coordinate $r$. As we said, acceptable or physical results have a critical point, $r=r_{c}$. That is, the fluid's velocity increases monotonically along its trajectory (at least until the event horizon is reached), and the flow is smooth in all points, which means that both the numerator and denominator in Eq. (\ref{Critical1}) vanish at the same point, namely, the critical point. Therefore, at the critical point one reads
\begin{equation}
u_{c}^{2}=\frac{r_{c}f'(r_{c})}{4}\ \ \ \mbox{and}\ \ \ V_{c}^{2}=\frac{r_{c}f'(r_{c})}{4f(r_{c})+r_{c}f'(r_{c})}.\label{u_Vc}
\end{equation}
For many
physically relevant cases, at the critical 
 point one has $u_c^{2}=c_s^{2}$, where
 \begin{equation}
 c_s=\sqrt{\frac{dp}{d\rho}}
 \end{equation}
 stands for the speed of sound in the   fluid. We will indeed use such condition here, but
 we need to keep in mind that this is not really
    obligatory, see \cite{Pacheco} for further references on this issue. 

 It is worth mentioning that for $r_{0}=0$ we can obtain the same results of Michel and the Schwarzschild case, i.e., $u_{c}^{2}=M_{0}/2r_{c}$ and $V_{c}^{2}=u_{c}^{2}/(1-3u_{c}^{2})$. Acceptable solutions require $u_c^2>0$ and $V_c^2>0$ at the critical point, therefore that implies some restrictions or constraints on the mass function and its parameters:
\begin{equation}
m'(r_c)-\frac{m(r_c)}{r_c}<0
\label{Inq1}
\end{equation}  
and
\begin{equation}
m'(r_c)+\frac{3m(r_c)}{r_c}-2<0.
\label{Inq2}
\end{equation}
The second restriction, for the Schwarzschild case,\footnote{There is a misprint in Michel's work \cite{Michel} concerning this point. In the quoted article, we read $r_c>6M_0$ instead of $r_c>3/2M_0$.} is simply $r_c>3/2M_0$. The inequalities (\ref{Inq1}) and (\ref{Inq2}) should impose a critical point outside the event horizon. Then the critical velocity, $u^2_c=c_s^2$, is reached outside the BH. 

The third equation for the accretion process is obtained from another conservation: $u_{\mu}T_{;\nu}^{\mu\nu}=0$. This provide us a continuity equation, that is to say, 
\begin{equation}
u^{\mu}\rho_{,\mu}+(p+\rho)u_{;\mu}^{\mu}=0,
\label{uT}
\end{equation}
in which the comma indicates an ordinary differentiation. By using the metric (\ref{Metric}), the above equation leads to
\begin{equation}
ur^{2}\exp\left(\int_{\rho_{\infty}}^{\rho}\frac{d\rho}{p+\rho}\right)=-A.
\label{A}
\end{equation}
Since $u<0$ for accretion, then $A$ must be positive. As we will see, the constant $A$ is related to the accretion rate of the perfect fluid. From (\ref{C2}) and (\ref{A}), one reads 
\begin{equation}
(p+\rho)(f(r)+u^{2})^{\frac{1}{2}}\exp\left(-\int_{\rho_{\infty}}^{\rho}\frac{d\rho}{p+\rho}\right)=-\frac{C_{2}}{A}=C_{4},
\label{C4}
\end{equation}
and $C_{4}=p_{\infty}+\rho_{\infty}$ (values calculated at infinity with fluid at rest, $u_{\infty}\rightarrow0$), assuming an asymptotically flat spacetime, like the Schwarzschild BH and the class of RBHs studied here. 

The
 change in the BH mass due the accretion of matter-energy  may be evaluated  \cite{Babichev1}  as 
\begin{equation} 
 \dot{M}=\oint T_{0}^{1}\sqrt{-g}d\theta d\phi ,
\end{equation} 
  where dot indicates time derivative, and $g$ is
 the metric determinant. For the metric given by Eq. (\ref{Metric}), we have
 simply  $\dot{M}=-4\pi r^{2}T_{0}^{1}$, and by using Eqs. (\ref{A}) and (\ref{C4}), we conclude that the mass variation is\footnote{Our constant $A$ differs from
  \cite{Babichev1,Babichev2} by a factor $M_0^{2}$.}
\begin{equation}
\dot{M}=4\pi A\left(p_{\infty}+\rho_{\infty}\right).
\label{DM}
\end{equation}
Therefore, the BH mass decreases with a phantom fluid because, as we said, $p+\rho<0$ for this type of perfect fluid. This is a quite interesting conclusion pointed out by Babichev \textit{et al.} \cite{Babichev1} with consequences   in cosmology, as we will see.  

The constants obtained in this section are important in order to calculate the radial velocity and the accretion rate, 
which are dependent on an EoS. In the next Section, we will use another EoS (instead of $p=w\rho$), which 
avoids both a nonsense speed of sound, in the case of negative values of $w$, and hydrodynamic instabilities when $\rho<0$.

\section{Critical points and other examples of fluids}

\begin{figure}
\begin{centering}
\includegraphics[scale=0.48]{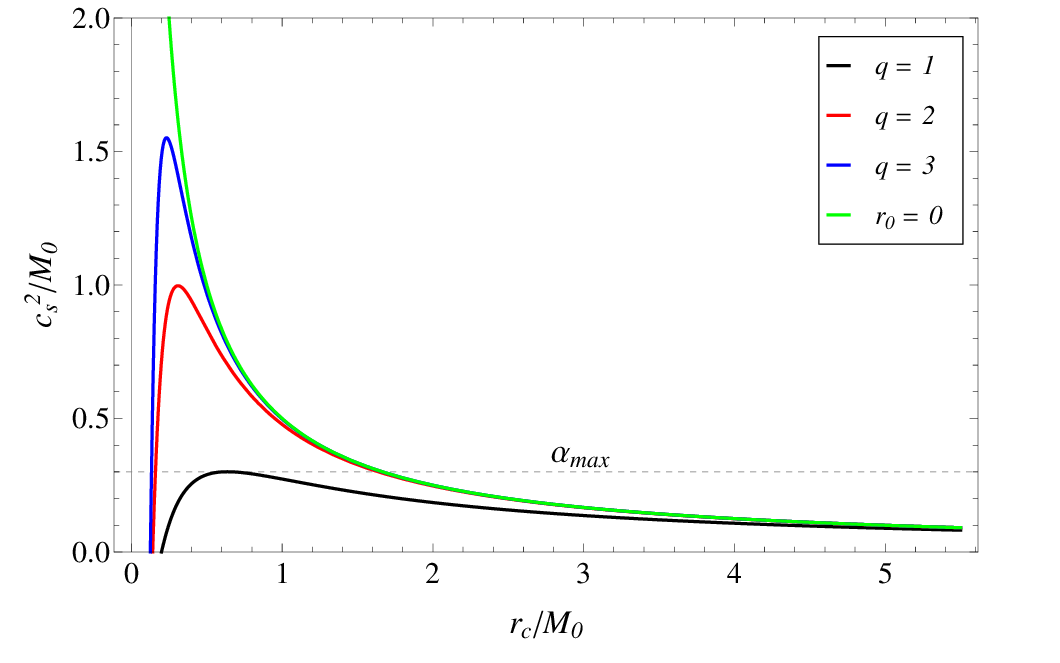}
\includegraphics[scale=0.48]{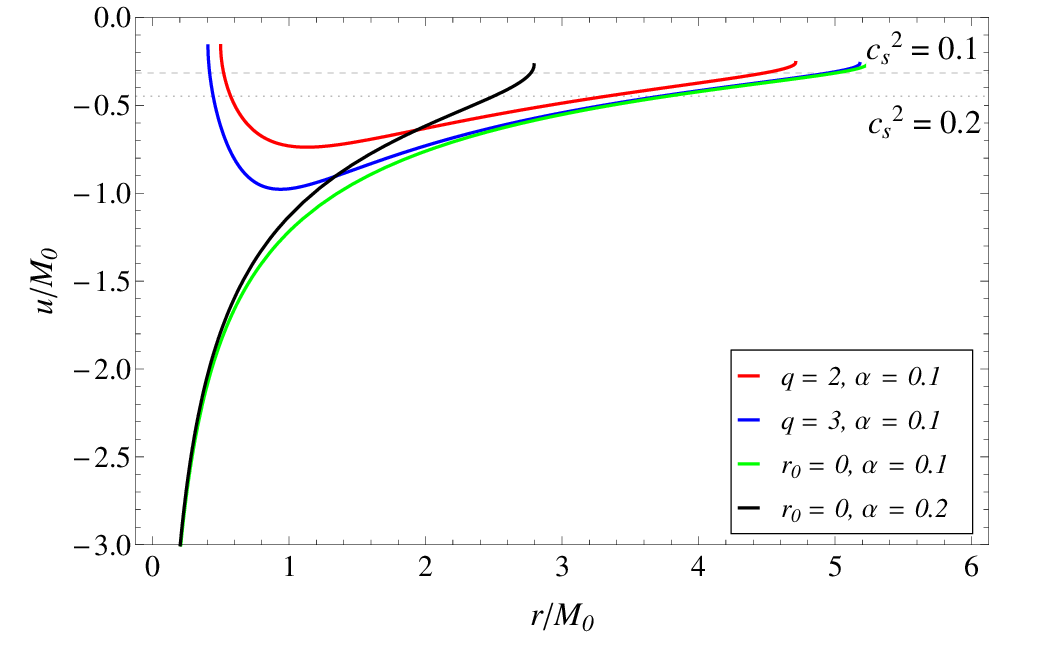}
\includegraphics[scale=0.47]{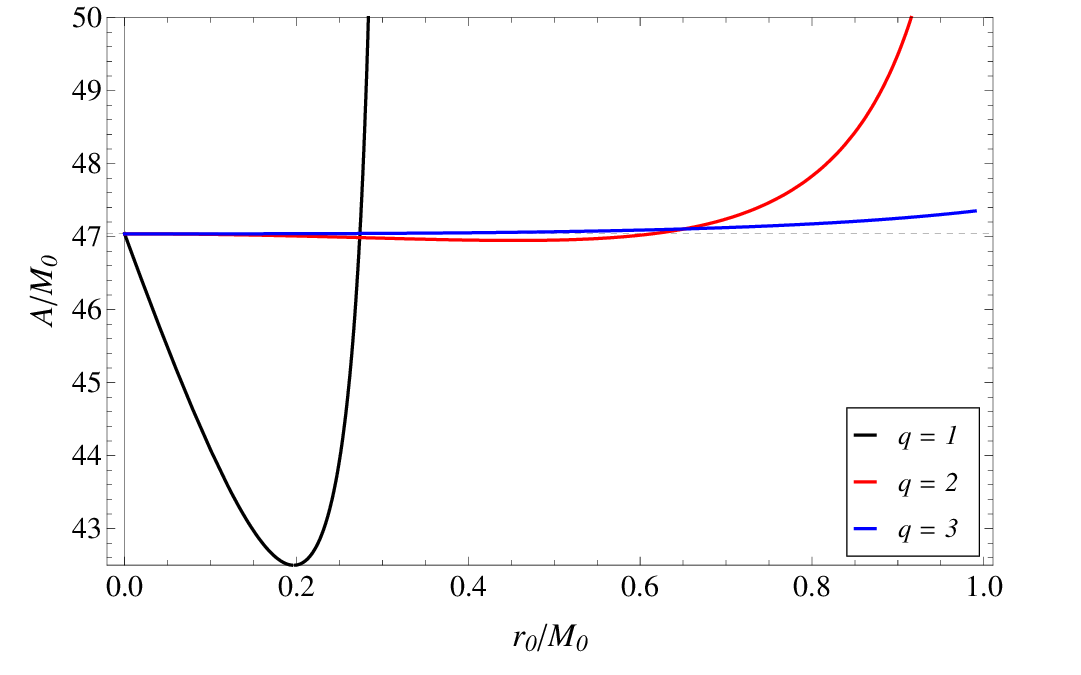}
\par\end{centering}

\caption{On the top, graphic for speed of sound over values of the critical point $r_c$, according to Eq. (\ref{Critcal_points}) and the equality $c_{s}^{2}=\alpha$. As we can see, there is a $\alpha_{max}$ in order to have a critical point outside the event horizon. In the middle, graphic of the radial velocity for some values of $q$ in Eq. (\ref{Mass_term}) and for the Schwarzschild black hole ($r_0=0$). It is interesting to note the fluid's behavior according to values of $\alpha$: the larger $\alpha$, the larger the speed of sound and, consequently, the critical point is smaller. That is to say, for large values of $\alpha$, the fluid in the accretion process needs more distance in order to reach $c_{s}$, the speed of sound. On the bottom, graphic for $A$, which provides the accretion rate, according to Eq. (\ref{DM}). The dashed line indicates the value of $A$ for the Schwarzschild BH. In these graphics, we used $M_0=1$ and $r_0=0.1$, on the top,  and $r_0=0.7$, in the middle.}
\label{Figure}
\end{figure}

In this section, a different EoS is adopted. Following Babichev \textit{et al.} \cite{Babichev1,Babichev2}, we assume that
\begin{equation}
p=\alpha(\rho-\rho_{0}),
\label{p}
\end{equation}
where $\rho_{0}$ is a constant. According to the cited authors, the above EoS avoids the hydrodynamic instabilities related to negative values of $\rho$. Another issue fixed by Eq. (\ref{p}) concerns the positivity of speed of sound for any type of fluid. For fluids with $w<0$, the EoS $p=w\rho$ leads to a nonsense result because $c_s=\sqrt{dp/d\rho}=w$. The EoS (\ref{p}) avoids such a problem because $\alpha$ may be adjusted in order to be positive. Then, from now, $c_s^2=u_c^2=\alpha$. It is worth mentioning that $\alpha$ has a maximum value in order to provide an accretion process in which the radial velocity is sonic outside the event horizon. As we can see in Fig. \ref{Figure} or by
using Eq. \ref{u_Vc} (with the conditions $u_{c}^2=\alpha$ and $V_{c}^2>0$, and $m_l(r)$, the approximation for 
the mass function), if $\alpha<\alpha_{max}$, we will have $r_c >r_+$. For the Schwarzschild BH, $\alpha_{max}=1/3$. Expressions for $\alpha_{max}$ in the class of RBHs are cumbersome, however for $q=3$, or the Hayward RBH, its approximate form is quite simple, namely 
\begin{equation}
\alpha_{max}\simeq \frac{27M_0^3-32r_0^3}{81M_0^3},
\end{equation}
 and the corresponding Schwarzschild limit   is straightforward. It is possible to relate both EoS (inasmuch as $w=\alpha(\rho-\rho_0)/\rho$) for some important fluids in cosmology, assuming $\rho>0$ and $0<\alpha<\alpha_{max}$:  
\begin{enumerate}
\item $w<-1$, phantom fluid: $-\frac{\rho}{\rho-\rho_0}<\alpha<\alpha_{max}$ with $\rho_0>0$ and $0<\rho<\rho_0$;
\item $w=-1$, cosmological constant: there is no accretion process;
\item $w=0$, dust: $\rho=\rho_0>0$;
\item $w=\frac{1}{3}$, radiation: $\alpha=\frac{\rho}{3(\rho-\rho_0)}$ and $\alpha_{max}>\frac{\rho}{3(\rho-\rho_0)}$ with $\rho_0 \leq 0$ and $\rho>0$ or $\rho_0>0$ and $\rho>\rho_0$;
\item $w=1$, stiff fluid: $\alpha=\frac{\rho}{\rho-\rho_0}$ and $\alpha_{max}>\frac{\rho}{\rho-\rho_0}$ with $\rho_0 \leq 0$ and $\rho>0$ or $\rho_0>0$ and $\rho>\rho_0$.
\end{enumerate}
With the choice of the EoS, physical quantities may be presented in terms of $\alpha$. From Eq. (\ref{A}), by using the EoS (\ref{p}), we have the following equation that relates thermodynamic and dynamic variables:
\begin{equation}
\left(\frac{p+\rho}{p_{\infty}+\rho_{\infty}}\right)=\left(-\frac{A}{ur^{2}}\right)^{1+\alpha}.
\label{P_+_rho}
\end{equation}
Moreover, with the aid of Eqs. (\ref{C2}) and (\ref{P_+_rho}), we obtain an easier expression in order to calculate the radial velocity:
\begin{equation}
f(r)+u^{2}=\left(\frac{p_{\infty}+\rho_{\infty}}{p+\rho}\right)^{\frac{2\alpha}{1+\alpha}}=\left(-\frac{ur^{2}}{A}\right)^{2\alpha}.
\label{u}
\end{equation}
As we can see in Fig \ref{Figure}, the radial velocity depends on $\alpha$. The larger $\alpha$, the closer to the event horizon the critical point is. That is, a sonic radial velocity for the fluid is reached just nearby $r_+$, then the perfect fluid (coming from infinity) takes more time in order to reach and surpass $c_s$ for large values of $\alpha$. 

With the EoS (\ref{p}) and the fact that $u_c^2=c_s^2=\alpha$ at the critical point, the constant $A$, given by Eq. (\ref{A}), is rewritten as
\begin{equation}
A=r_{c}^{2}\left(\frac{\alpha^{\alpha}}{\alpha+f(r_c)}\right)^{\frac{1}{2\alpha}}.
\label{A2}
\end{equation}
By assuming the positivity of $A$ for any value of $\alpha$ in order to obtain a general accretion process, 
the denominator of Eq. (\ref{A2}) must be positive, and it will be if $\alpha+f(r_c)>0$, which is translated 
into Eq. (\ref{Inq2}). Moreover, for large values of $r$ with the aid of (\ref{Mass_approx_large}), we can see 
an easier expression for $A$ and compare to the Schwarzschild BH, that is to say,
\begin{equation}
A \simeq r_c^2 \left(\frac{\alpha^\alpha}{\left(\alpha+1-\frac{2M_0}{r_c}+\frac{6M_0 r_0^q}{q r_c^{q+1}}  \right)}\right)^{\frac{1}{2\alpha}}.
\label{A_approx}
\end{equation}  
For some values of $q$ and $r_0$ in Eq. (\ref{A_approx}), the term that contains those parameters may produce a smaller  accretion process onto RBHs compared to the Schwarzschild BH (see Fig. 1 for comparison). With $r_0\neq0$, 
the denominator of (\ref{A_approx}) may be larger (for $q=1$ and $q=2$), then, according to Eq. (\ref{DM}), the accretion process takes more time for RBHs, that is to say, their accretion rates are smaller compared to the 
Schwarzschild BH. Differences in $A$ are due to different values assumed by $q$, $r_0$, and  $r_c$, which 
depends on that parameters. For the entire class of RBHs, the critical point $r_c$ is smaller than the critical 
point of the Schwarzschild metric. And in the cases in which $q=1$ and $q=2$, considering some values of $r_0$,  
the constant $A$ is smaller as well. On the other hand, the accretion rate is larger for different values of $q$, 
suggesting a larger accretion process even compared to the process onto the Schwarzschild BH. With a larger
$A$, the radial velocity is larger as well, according to Eq. (\ref{A}). For those values of $A$,  RBHs enhance
the accretion process instead of decreasing it.

For $r=r_c$, Eq. (\ref{u}) (left and right sides) provides the equation whose roots are the critical points. From Eqs. (\ref{u}) and (\ref{A2}), by using both the metric and the mass function, the critical points are solutions of
\begin{equation}
\frac{M_0}{2r_c}\left(1-2\left(\frac{r_{0}}{r_{c}}\right)^{q}\right)\left(1+\left(\frac{r_{0}}{r_{c}}\right)^{q}\right)^{-\frac{3+q}{q}}=\alpha.
\label{Critcal_points}
\end{equation}
Another way to obtain the critical points and the above equation is simply solving $u_c^2=\alpha$, with the aid of Eq. (\ref{u_Vc}). In general, for $r_{0}\ll M_{0}$ and any positive integer $q$, Eq. (\ref{Critcal_points}) presents two positive solutions or critical points: the inner $r_{c-}$ and the outer critical point $r_{c+}$. These points, which are solutions of (\ref{Critcal_points}), obey the following relation:
\begin{equation}
0<r_-<r_{c-}<r_+<r_{c+}.
\end{equation}
For the Schwarzschild BH, there is just one critical point: $r_{c+}=M_0/2 \alpha$. As we said, $r_c=r_{c+}>3M_0/2$, thus $0< \alpha< 1/3$. That is the origin for $\alpha_{max}=1/3$ in the Schwarzschild spacetime as we already pointed out. And, according to Fig. \ref{Figure}, the class of RBHs also presents a maximum $\alpha_{max}$, which is about $1/3$ (as close as $r_0 \approx 0$).

The role played by $r_0$ is important in order to obtain the critical points. Following Ref. \cite{Maluf_Neves1}, where a GUP was applied to the RBHs studied here, the same GUP was used in the Bardeen metric in Ref. \cite{Maluf_Neves2}, then the length parameter was obtained and written as
\begin{equation}
r_0=\frac{\lambda l_p}{3},
\label{r0}
\end{equation}
in which $\lambda$ stands for the dimensionless quantum gravity parameter, and $l_p$ is the Planck length. When
 $\lambda=0$ in the generalized principle, we recover the Heisenberg uncertainty relation, and, according to several
 methods, $\lambda$ (and consequently $r_0$) may be estimated. In \cite{Das_Vagenas,Scardigli_Casadio}, some
methods in order to obtain the quantum gravity parameter are commented. The best upper bound for $\lambda$ is
obtained from the scanning tunneling microscope, that is to say, $\lambda<10^{10}$ for that method. Thus, $r_0<10^{-25}$m is straightforwardly obtained from Eq. (\ref{r0}), and Table \ref{Values of rc} presents some values of $r_{c+}$ estimated from some geometries and for perfect fluids with EoS $\alpha=0.1$ and $\alpha=0.2$ in the accretion processes of objects with  one solar mass and $\sim10^{6}$ solar masses, like Sagittarius A*, the super massive BH at the Milk Way center. Such as Fig. \ref{Figure} indicates, large values of $\alpha$ produce small values of the critical point, showing that the fluid in the accretion process needs more distances in order to acquire a radial velocity equal to speed of sound. Above all, the tiny value of $r_0$ imposes that the difference of critical points among members of the class of RBHs and the Schwarzschild BH is pretty much indistinguishable for astrophysical BHs. As we said, such a microscopical value becomes very important just at small scales, inside the event horizon. However, if clear differences between RBHs and singular BHs are not noticeable regarding large distances, it will be over a long time. In this sense, a cosmological scenario may provide us some differences. 

 \begin{table*}
\caption{Values of the outer critical point, $r_c=r_{c+}$, for the Schwarzschild BH and members of the class of RBHs
 by using perfect fluids with EoS $\alpha=0.1$ and $\alpha=0.2$ in order to describe the accretion processes for objects with one solar mass ($M_{\odot}$) and $10^{6}$ solar masses like the Sagittarius A*. All critical point values for RBHs are obtained using the value for the quantum gravity parameter that produces $r_0\sim10^{-25}$m, estimated in
 \cite{Maluf_Neves2}. Such a value for $r_0$ turns indistinguishable BHs and RBHs critical points from that point of view. As we can see, the larger $\alpha$, from EoS (\ref{p}), the closer the critical point is. Keep in mind that the event horizon for those objects is about 3 km for $M_0=M_{\odot}$ and $1.2\times10^{7}$ km for $M_0=4 \times 10^{6}M_{\odot}$. }
\label{Values of rc}
\begin{ruledtabular}
\begin{tabular}{ccccccc}
     &&& \text{Critical Point $r_c=r_{c+}$\footnote{Values in kilometers.}}&&& \\ \hline
     
     && $\alpha=0.1$              &&             & $\alpha=0.2$&          \\
              
    \text{Type of Black Hole} & $M_0=M_{\odot}$ &&  $M_0=4\times10^{6}M_{\odot}$ & $M_0=M_{\odot}$ && $M_0=4\times10^{6}M_{\odot}$ \\ \hline
    
     \text{Schwarzschild BH and RBHs} & $\simeq7.428$ && $\simeq2.986 \times 10^{7}$  & $\simeq3.714$ && $\simeq1.493 \times 10^{7}$\\ 
           
\end{tabular}
\end{ruledtabular}
\end{table*}

\section{Implications for cosmology}
The accretion process onto black holes may be an important issue in cosmology. If, as we saw, a cosmological constant
conceived of as a perfect fluid cannot be accreted by BHs and RBHs, other types of fluids will be. As we saw, according to \cite{Babichev1}, a phantom fluid may decrease the BH mass during the accretion process. Then, somehow, destruction of
  BHs and RBHs by means of phantom energy turns into an issue in bouncing and cyclic cosmologies. According to cyclic
   cosmologies  \cite{Brown,Frampton}, the initial singularity and the big rip are avoided, and before the bounce and after the turnaround, a contraction  phase  is predicted. The scale factor reaches minimum and maximum values at the bounce and the turnaround, respectively. Whether before the bounce or the turnaround, BHs should be eliminated in order to promote   
a  new cycle   without such debris. According to the authors of \cite{Brown}, BHs are torn apart in the expansion
       phase---dominated by phantom energy---before the turnaround. But in Ref. \cite{Sun}, this very conclusion is criticized. Accordingly, BHs do  not evaporate before the turnaround using a phantom fluid. Their masses decrease during
        the late expansion dominated  by a phantom fluid and,  then, the remaining masses are just eliminated during the
         contraction phase by the Hawking
   radiation. Following  Ref. \cite{Sun}, who adopts a modified Friedmann equation in order to provide a turnaround,\footnote{A modified Friedmann equation that generates cyclic cosmologies is found in contexts beyond general relativity like the brane world \cite{Brown}.} the BH mass at the turnaround for initial masses larger than the Planck mass is 
\begin{equation}
M_c\simeq \sqrt{\frac{3}{2\pi^3\rho_c}}\frac{M_p^3}{A },
\label{Critical_mass}
\end{equation}    
in which $\rho_c$, according to the author, is the phantom's critical density at the turnaround, and $M_p \sim 10^{-8}$ kg is the Planck mass. Here, the initial mass is $M_0$, which is given at the beginning of the phantom dominated phase. Therefore, using Eq. (\ref{A_approx}) in order to compare accretion processes onto BHs and
 RBHs, we conclude that the RBH mass may be larger than the Schwarzschild mass at the turnaround in a cyclic cosmology, especially for $q=1$ and $q=2$. Such a result is in agreement with the previous conclusion of ours, according to which the accretion process is slower for    RBHs with  $q=1$ or $q=2$. Thus, a phantom dominated phase will steal more mass from the Schwarzschild BH  than from that RBHs. Moreover, as we said, according to Ref. \cite{Sun}, the remaining mass, whether of BH or RBH, is evaporated by Hawking's mechanism in the contraction phase, and a phantom cyclic cosmology is safe from that astrophysical objects in this perspective.

In order to promote another direct comparison between BHs and RBHs, we propose a critical mass ratio at the turnaround for those objects, that is to say, by using the approximate $A$, given by Eq. (\ref{A_approx}), and the critical mass defined above, we have
\begin{equation}
\frac{M_{c(RBH)}}{M_{c(Sch)}}\simeq \left(1+\mathcal{F}(\alpha,r_0,M_0,q) \right)^\frac{1}{2\alpha},
\label{Mass_ratio}
\end{equation}
where $M_{c(RBH)}$ and $M_{c(Sch)}$ stand for the critical mass for RBHs and the Schwarzschild BH at the turnaround, respectively, and
\begin{equation}
\mathcal{F}(\alpha,r_0,M_0,q)=\frac{6M_0}{\left(\left(1+\alpha \right)r_c-2M_0 \right)q}\left(\frac{r_0}{r_c}\right)^q
\end{equation} 
is a positive function. Following \cite{Sun}, the critical mass $M_c$ is about the Planck mass at the turnaround. With a
 tiny quantum gravity parameter that provides $r_0$, the function $\mathcal{F}(\alpha,r_0,M_0,q)$ renders an
  indistinguishable difference between astrophysical BHs and RBHs at the turnaround (astrophysical at the beginning of the
   phantom dominated era). That is to say, by assuming $M_0\sim M_{\odot}$ or $M_0\sim 10^{6}\times M_{\odot}$ as
    the initial masses for the Schwarzschild BH and RBHs at the beginning of the phantom dominated phase, the difference
     between singular and RBHs is still very tiny at the turnaround. However, if the initial masses are similar to the
      primordial black hole (PBH) masses, it will be more interesting. PBHs are hypothetical objects generated during 
       the radiation dominated era and are candidates for dark matter. Recently \cite{PBH}, the Subaru Telescope constrained the PBHs masses range to $10^{-11}M_{\odot} - 10^{-6}M_{\odot}$ by using the microlensing effect of stars in M31
        (Andromeda Galaxy) caused by PBHs in the halo of the Milky Way. According to \cite{PBH}, with these masses,
         PBHs as candidates for the total amount of dark matter are pretty much ruled out.  However, with $M_0\simeq 10^{-11} M_{\odot}$ as the initial mass during the phantom era, Eq. (\ref{Mass_ratio}) with $q=1$ and $\alpha = 0.2$ renders about $10^{-19}$ kg of difference between the Schwarzschild BH and RBHs. That is, both objects will have
          approximately Planck masses at the turnaround, but RBHs will have a tiny advantage in their masses ($\simeq 10^{-19}$ kg). Then, within a cosmological point of view, it is possible to conceived of a quantitative difference  between the Schwarzschild BH and RBHs.

\section{Final remarks}
We applied Michel's \cite{Michel} approach, which describes steady-state accretion onto BHs, to a class of  RBHs developed in Ref. \cite{Neves_Saa}. Compared to the Schwarzschild BH, some members of that class present a potentially slower accretion process. Regarding the accretion process of a phantom fluid, the RBHs masses may decrease with a slower rate compared to Schwarzschild's. The origin for that is found in the accretion rate and the radial velocity of the accreted perfect fluid. For some members of the class of RBHs, the radial velocity and the accretion rate are smaller
than the radial velocity and accretion rate of the Schwarzschild BH. For dust as a perfect
fluid, a slower radial velocity for the entire class of RBHs indicates the capability of the de Sitter core---that 
which generates regular solutions---of playing the role of an antigravity source.  

By using the recent upper bound on the parameter that provides regular metrics, namely $r_0<10^{-25}$m according to \cite{Maluf_Neves2}, we calculated the critical point, where the fluid's radial velocity is equal to speed of sound in the fluid. Due to the tiny value of $r_0$, members of the class of RBHs present indistinguishable values for the critical points compared to Schwarzschild BH.

However, from the cosmological point of view, differences between RBHs and the Schwar-\ zschild BH are suggested. With a slower accretion process for some RBHs, the evaporation of RBHs due to a phantom fluid takes longer time, in agreement with other results. Therefore, the accretion process conceived over long time may produce different masses between BHs and RBHs. In particular, RBHs with $q=1$ and $q=2$ would have larger masses than the Schwarzschild BH at the turnaround in cyclic cosmologies. This result may be important for cyclic cosmologies, in which the existence of debris of previous cycles are avoided in order to build realistic models.

\section*{Acknowledgments}
The   authors acknowledge the financial support of CAPES (Coordenação de Aperfeiçoamento de Pessoal de Nível
 Superior, Finance Code 001), CNPq, and FAPESP (Grant 2013/09357-9).  JCSN thanks Vilson Zanchin's research
  group for comments and suggestions during a seminar presentation at the UFABC, and AS is grateful for the warm hospitality at the Department of Theoretical Physics,
   University of Zaragoza, where part of this paper was conducted. We would
like to thank Ronaldo S.S. Vieira for reading the manuscript. We also thank an anonymous referee for 
comments and suggestions.

\end{document}